%% file: main.tex
\documentstyle[seceq]{ptptex}
\newcommand{\dslash}{\ooalign{\hfil/\hfil\crcr$\partial$}}

\notypesetlogo
\begin{document}
\title{Bosonic Structure of a 2-Dimensional Fermion Model\\
with Interaction among Different Species II\\
-- $N$-species Case --}
\author{Jiro {\sc Sakamoto}\thanks{E-mail
address:\ jsakamot@riko.shimane-u.ac.jp}\ \ and Yasuaki {\sc Fukuoka}}
\inst{Physics department, Shimane University, Matsue 690-8504, Japan}
\abst{\input{abstract}}
\maketitle
\input{intro}

\setcounter{equation}{0}
\input{model}
\setcounter{equation}{0}
\input{perturb}

\setcounter{equation}{0}
\input{discussion}

\setcounter{equation}{0}

\input{references}
\end{document}

%% file: abstract.tex
We study a massive Thirring-like model in 2-dimensional space-time,
which contains fermions with arbitrary number ($N$) of different
species. This model is an extension of that of a previous paper, where we 
have considered two-species case. By this extension we expect that we
can expose more general structures of this kind of model. We obtain the
equivalent boson model with $N$ species to our fermion model. We find
that the coupling constant must be set in some regions in order for the model
to be physically sensible. It seems hard to find such regions
from  direct obsavation of the original fermion model. We also find that
for specific values of the coupling constant some of the boson fields disappear from the system. Therefore, the $N$-species fermion model
is described by the boson model with fewer species. 
 

%% file: intro.tex
\section{Introduction}

The bosonization
technique is one of the powerful approaches to study a
2-dimensional fermion system.\cite{bosonization} As Coleman has done in
his pioneering work,\cite{Coleman} with this technique one can expose
hidden properties of such a fermion system though it may be
applicable only for charge-zero sectors of fermions, i.e. for sectors of
pairs of fermion and anti-fermion. In a previous paper\cite{previous}
(I) using the path-integral method we have derived an equivalent boson
model to a two-species fermion model with interaction only among different
species. We have found that in order for the model to be physically
sensible the coupling constant must be set in some regions. We alse
found that for a specific value of the coupling constant  $g=2\pi/3$ one
boson is decoupled from the system and that the two-species fermion
system can be described by the one-species boson model, i.e. the
ordinary sine-Gordon model.   

In this article we consider the bosonization of the $N$-species fermion
model, which is a direct extension of the model in I. It is an easier
but non-trivial extension of the freedom of the bosonization technique.
By this extension  we expect that we can expose more general structures
of this model.  As is shown below, we derive the $N$-species equivalent
boson model and  find that the coupling constant
$g$ must be in the regions $g< -\pi/(N-1)$, $-2\pi/3(N-1)\leq g \leq
2\pi/3$ or $g> \pi$ in order for this boson model to be physically
sensible. The region corresponding to the second one was missed in I. We
also find that for specific values of the coupling constant some bosons
are decoupled from the system, i.e. one or $N-1$ bosons are decoupled
for $g=-2\pi/3(N-1)$ or $g=2\pi/3$, respectively. Therefore, the
$N$-species fermion model is equivalent to $(N-1)$-species boson model
for the former case and to one-species boson model for the latter case. 

We use the same notations as in I, i.e. in Minkowski space-time,
$g_{\mu\nu}=(-1,+1)$ and
$\epsilon^{01}=-\epsilon_{01}=1$. Gamma-matrices are given as
$\gamma^{0}=-\gamma_0=i\sigma_x$, $\gamma^1=\gamma_1=\sigma_y$,
$\gamma_5=\gamma_0\gamma_1=\sigma_z$, where $\sigma_x, \sigma_y$ and
$\sigma_z$ are the Pauli matrices.

%% file: model.tex
\section{Model}
Our initial Lagrangian is given by
\begin{equation}
 {\cal L}= 
\sum_{i=1}^{N}\overline{\psi}_i({\ooalign{\hfil/\hfil\crcr$\partial$}}-m)\psi_i+\sum_{i>j=1}^{N}{1\over 2}gj_{i\mu}j_j^{\mu},\label{lagrangian1}
\end{equation}
where $i, j$ denote the fermion species and 
\begin{equation}
 j_{i\mu }= i\overline{\psi}_{i}\gamma_{\mu}\psi_{i}.\label{current_i}
\end{equation}
In (\ref{lagrangian1}) a fermion  interacts with those  of the
different species and never with itself  directly. When $N=2$, we can
change the sign of $g$ by taking charge conjugation of one of the fermion
species. In fact we saw in I that the consequences for $N=2$ are symmetric for
$g\leftrightarrow -g$. On the contrary, for the case $N\geq 3$, there is
not such a symmetry, and therefore the sign of $g$ has a physical meaning
for this case. 

The quartic interaction part in (\ref{lagrangian1}) is rewritten as

\begin{equation}
 \sum_{i>j=1}^{N}{1\over 2}gj_{i\mu}j_j^{\mu} = \frac{g}{4}\left(\sum_i j_{i\mu}\right)^2 - \frac{g}{4}\sum_i j_{i\mu}j_{i}^{\mu},
\end{equation}
and this is equivalent to 
\begin{equation}
 \frac{g}{2}\sum_i j_{i\mu}X^{\mu}- \frac{g}{4}X_{\mu}X^{\mu}+\frac{g}{2}\sum_i\left(\frac{1}{2}A_{i\mu}-j_{i\mu}\right)A_i^{\mu},
\end{equation}
with using auxiliary vector fields $X_{\mu}$ and $A_{i\mu}$. Now we put
\begin{equation}
 B_{i\mu}=X_{\mu}-A_{i\mu},
\end{equation}
and integrate out over $X_{\mu}$ by the path-integral formulation to
obtain
\begin{equation}
 \frac{g}{2}\sum_i j_{i\mu}B_i^{\mu}- \frac{g}{4(N-1)}\left(\sum_i B_{i\mu}\right)^2 + \frac{g}{4}\sum_i\left(B_{i\mu}\right)^2.
\end{equation} 
Here we should note that for $N\geq 3$ the above expression contains
direct coupling among the same species of boson fields $B_{i\mu}$ while
fermion fields do not interact directly with themselves in the original
Lagrangian (\ref{lagrangian1}).

In 2-dimensional space-time, we can write  vector fields $B_{i\mu}$ with
two scalar fields $\phi_i$ and $\chi_i$ as
\begin{equation}
 B_{i\mu}=\epsilon_{\mu\nu}\partial^{\nu}\phi_i + \partial_{\mu}\chi_i.
\end{equation}
With these scalar fields we transform the fermion fields as
\begin{equation}
 \psi_i\rightarrow \psi_i'=\exp\left\{\frac{ig}{2}(-\gamma_5\phi_i+\chi_i)\right\}\psi_i,
\end{equation}
to rewrite the Lagrangian as
\begin{eqnarray}
 {\cal L}_{\rm eff}&=&\sum_i \overline{\psi}_i'(\dslash -m\exp\{ig\gamma_5\phi_i
\})\psi_i' - \frac{g}{4(N-1)}\left\{\left(\sum_i\partial_{\mu}\chi_i\right)^2-
\left(\sum_i\partial_{\mu}\phi_i\right)^2\right\}
\nonumber \\
&&+\frac{g}{4}\sum_i\left(\partial_{\mu}\chi_i\right)^2 
+\left(\frac{g^2}{4\pi}-\frac{g}{4}\right)\sum_i\left(\partial_{\mu}\phi_i\right)^2. 
\end{eqnarray}
In the above expression the term $(g^2/4\pi)(\partial\phi)^2$ comes from
det$|\exp(-ig\gamma_5\phi_i)|$ in the path-integral measure following Fujikawa.\cite{Fujikawa}
It is seen that $\chi_i$ is decoupled from the other fields and  can be
integrated out. Then we obtain
\begin{eqnarray}
 {\cal L}_{\rm eff} &=& \sum_i \overline{\psi}_i\left(\dslash -m\exp\{ig\gamma_5\phi_i\}\right)\psi_i \nonumber \\
&+& \frac{g}{4(N-1)}\left(\sum_i\partial_{\mu}\phi_i\right)^2 
+ \frac{g}{4}(\frac{g}{\pi}-1)\sum_i\left(\partial_{\mu}\phi_i\right)^2,\label{L_eff}
\end{eqnarray}
where we write $\psi_i$ for $\psi_i'$. 

The generating functional of the Green functions is given by
\begin{equation}
 Z = \int \prod_i d\overline{\psi}_i d\psi_i d\phi_i \exp\left\{i\int d^2x {\cal L}_{\rm eff}\right\}.\label{Z}
\end{equation}
By the `Wick' rotation we transform ourselves into  Euclidean space-time
from the Minkowski space-time. Then the generating functional (\ref{Z}) is 
rewritten in Euclidean space-time as
\begin{equation}
 Z_{\rm E}=\int\prod_i d\overline{\psi}_id\psi_id\phi_i \exp\{-\int d^2x{\cal L}_{\rm eff}\}. \label{Z_E}
\end{equation}

%% file: perturb.tex
\section{Perturbative expansion}
In order to calculate (\ref{Z_E}) by perturbation theory we choose
${\cal L}_0={\cal L}_{\bf eff}(m=0)$ of (\ref{L_eff}) as the free Lagrangian.  We
find that the free fermion and boson propagators are given as
\begin{eqnarray}
 \left\langle\overline{\psi}_i(x)\psi_j(y)\right\rangle&=&\frac{\delta_{i,j}}{2\pi}\frac{\gamma\cdot (x-y)}{(x-y)^2},\label{f_prop}\\
\left\langle\phi_i(x)\phi_j(y)\right\rangle&=&\frac{1}{2g}\left\{\frac{\pi}{(g-\pi)\{\pi+g(N-1)\}}-\frac{\delta_{i,j}}{g-\pi}\right\}\ln (x-y)^2\mu^2,\label{b_prop}
\end{eqnarray}
where  $<\cdots>$ denotes the vacuum expectation value of the
time-ordered product. Parameter $\mu$ is a small infrared cutoff mass,
which will be set to zero after the calculations.

Now, we calculate $Z_{\rm E}$ of (\ref{Z_E}) through the perturbative expansion
with respect to $m$. We note that all the odd order terms vanish because 
of traceless property of $\gamma$-matrices and the super selection rule
for the boson field $\phi$, i.e. $<\exp i\sum\beta_i\phi>=0$ unless
$\sum\beta_i=0$ where $\beta_i=\pm g$.\cite{bosonization,Coleman} The $2n$-th
order term of the expansion of $Z_E$ is given as
\begin{eqnarray}
 Z_{\rm E}^{(2n)}&=& \frac{m^{2n}}{(2n)!}\int\prod d\overline{\psi}_id\psi_id\phi_i
\left\{\int dx\sum_{i=1}^N\overline{\psi}_ie^{ig\gamma_5\phi_i}\psi_i\right\}^{2n}\exp\left\{-\int dx {\cal L}_{0}\right\}\nonumber \\
&= &\frac{m^{2n}}{(2n)!} \left\langle \left\{\int dx\sum_{i=1}^N\overline{\psi}_i
e^{ig\gamma_5\phi_i}\psi_i\right\}^{2n}\right\rangle.\label{Z_2n}
\end{eqnarray}
Using the identity
\begin{equation}
 \overline{\psi}e^{ig\gamma_5\phi}\psi = e^{ig\phi}\overline{\psi}\Gamma_+\psi + e^{-ig\phi}\overline{\psi}\Gamma_-\psi,
\end{equation}
where we put $\Gamma_{\pm}=(1\pm\gamma_5)/2$, we expand (\ref{Z_2n}) as
\begin{eqnarray}
 Z_{\rm E}^{(2n)}&=& m^{2n}\sum_{r_1,r_2,\cdots r_N}^n\frac{\delta_{n,r_1+r_2+\cdots +r_N}}{(2r_1)!(2r_2)!\cdots(2r_N)!}\nonumber\\
&&\times\left\langle\left\{\int dx\left(e^{ig\phi_1}\overline{\psi}_1\Gamma_+\psi_1+e^{-ig\phi_1}\overline{\psi}_1\Gamma_-\psi_1\right)\right\}^{2r_1}\right.\nonumber\\
&&\times \left\{\int dx\left(e^{ig\phi_2}\overline{\psi}_2\Gamma_+\psi_2 + e^{-ig\phi_2}\overline{\psi}_2\Gamma_-\psi_2\right)\right\}^{2r_2}\nonumber\\[10pt]
&&\times\cdots\cdots\cdots\cdots\cdots\nonumber\\[10pt]
&&\times\left.\left\{\int dx\left(e^{ig\phi_N}\overline{\psi}_N\Gamma_+\psi_N + e^{-ig\phi_N}\overline{\psi}_N\Gamma_-\psi_N\right)\right\}^{2r_N}\right\rangle\nonumber\\
&=& m^{2n}\sum_{r_1,r_2,\cdots r_N}^n\frac{\delta_{n,r_1+r_2+\cdots +r_N}}{(r_1!)^2(r_2!)^2\cdots (r_N!)^2}\int\prod dxdy\nonumber\\
&&\times\left\langle\prod_{i=1}^{r_1}\overline{\psi}_1(x_{1,i})\Gamma_+\psi_1(x_{1,i})\overline{\psi}_1(y_{1,i})\Gamma_-\psi_1(y_{1,i})\right\rangle\nonumber\\
&&\times\left\langle\prod_{i=1}^{r_2}\overline{\psi}_2(x_{2,i})\Gamma_+\psi_2(x_{2,i})\overline{\psi}_2(y_{2,i})\Gamma_-\psi_2(y_{2,i})\right\rangle\nonumber\\[10pt]
&& \cdots\cdots\cdots\cdots\nonumber\\[10pt]
&&\times\left\langle\exp ig\left[\sum_{i=1}^{r_1}\{\phi_1(x_{1,i})-\phi_1(y_{1,i})\}+\sum_{i=1}^{r_2}\{\phi_2(x_{2,i}-\phi_2(y_{2,i}))\}\right.\right.\nonumber\\
&&\left.\left.+\cdots\cdots\cdots 
+\sum_{i=1}^{r_N}\{\phi_N(x_{N,i})-\phi_N(y_{N,i})\}\right]\right\rangle.\label{Z_E^2n}
\end{eqnarray}
With the fermion propagator (\ref{f_prop}) we can calculate a fermion
part in the above (\ref{Z_E^2n}) as 
\begin{equation}
 \left\langle\prod_{i=1}^{r_k}\overline{\psi}_k(x_{k,i})\Gamma_+\psi_k(x_{k,i})\overline{\psi}_k(y_{k,i})\Gamma_-\psi_k(y_{k,i})\right\rangle= \frac{1}{(2\pi)^{2r_k}}\frac{\prod_{i>j}^{r_k}(x_{k,i}-x_{k,j})^2(y_{k,i}-y_{k,j})^2}{\prod_{i,j}^{r_k}(x_{k,i}-y_{k,j})^2},
\end{equation}
and we obtain
\begin{eqnarray}
 Z_E^{(2n)}&=& \left(\frac{m}{2\pi}\right)^{2n}\sum_{r_1,r_2,\cdots r_N}^{n}
\frac{\delta_{n,r_1+r_2+\cdots +r_N}}{(r_1!)^2(r_2!)^2\cdots (r_N!)^2}\int\prod dxdy\nonumber\\
&&\times\frac{\prod_{i>j}^{r_1}(x_{1,i}-x_{1,j})^2(y_{1,i}-y_{1,j})^2}
{\prod_{i,j}^{r_1}(x_{1,i}-y_{1,j})^2}\times\frac{\prod_{i>j}^{r_2}(x_{2,i}-x_{2,j})^2(y_{2,i}-y_{2,j})^2 }{\prod_{i,j}^{r_2}(x_{2,i}-y_{2,j})^2}\nonumber\\
&&\times\cdots\cdots\times \frac{\prod_{i>j}^{r_N}(x_{N,i}-x_{N,j})^2(y_{N,i}-y_{N,j})^2}{\prod_{i,j}^{r_N}(x_{N,i}-y_{N,j})^2}\nonumber
\end{eqnarray}
\begin{equation}
\times \left\langle \exp ig\left[\sum_{i=1}^{r_1}\left\{\phi_1(x_{1,i})-\phi_1(y_{1,i})\right\}+\sum_{i=1}^{r_2}\left\{\phi_2(x_{2,i})-\phi_2(y_{2,i})\right\}+\cdots \cdots \right]\right\rangle.\label{Z_2n_cal}
\end{equation}
Now, let us calculate the contractions among the boson fields with the
same species  in the above expression to obtain
\begin{eqnarray}
&& \left\langle\exp ig\left[\sum_{i=1}^{r_k}\{\phi_k(x_{k,i})-\phi_k(y_{k,i})\}\right]\right\rangle = \exp\left\{-\frac{g^2 r_k}{2}<\phi_k(0)^2>\right\}\nonumber\\
&&\times\exp -g^2\left[\sum_{i>j}^{r_k}\left\{<\phi_k(x_{k,i})\phi_k(x_{k,j})>+<\phi_k(y_{k,i})\phi_k(y_{k,j})>\right\}-\sum_{i,j}^{r_k}<\phi_k(x_{k,i})\phi_k(y_{k,j})>\right]\nonumber\\
&& = \exp\left\{-\frac{g^2r_k}{2}<\phi_k(0)^2>\right\}\nonumber\\
&&\times\exp\left\{\frac{g^2(N-1)}{2(g-\pi)\{\pi+g(N-1)\}}\ln\left(\frac{\prod_{i>j}^{r_k}(x_{k,i}-x_{k,j})^2\mu^2(y_{k,i}-y_{k,j})^2\mu^2}{\prod_{i,j}^{r_k}(x_{k,i}-y_{k,j})^2\mu^2}\right)\right\},
\end{eqnarray}
where we have used (\ref{b_prop}). Here we suppose that for a small
distance the boson propagator is properly regularized.
We find that each fermion part in (\ref{Z_2n_cal}) has the same
combination of the coordinates \{$x_{k,i}, y_{k,i}$\} as that in 
the above expression.

We, therefore, suppose that in the corresponding boson model the free
boson propagator gives such a fermion contribution besides the original
boson one (\ref{b_prop}) as 
\begin{eqnarray}
&& <\phi_i'(x)\phi_j'(y)>= <\phi_i(x)\phi_j(y)>- \frac{\delta_{i,j}}{g^2}\ln (x-y)^2\mu^2\nonumber\\
&&\ \ \ \ \ \ \ \  = \frac{1}{2g}\left\{\frac{\pi}{(g-\pi)\{\pi+g(N-1)\}}-\frac{3g-2\pi}{g(g-\pi)}\delta_{i,j}\right\}\ln (x-y)^2\mu^2,\label{b'_prop}
\end{eqnarray}
and that potential term is given by
\begin{equation}
 m'^2\left\{\cos g\phi_1' + \cos g\phi_2' + \cdots + \cos g\phi_N'\right\}.\label{b_potential}
\end{equation}
The $2n$-th order term of the perturbative expansion of the generating
functional for such a model is calculated as
\begin{eqnarray}
 Z_{\rm B}^{(2n)}&=& \frac{m'^{4n}}{(2n)!}\left\langle\left\{\int dx\left(\cos g\phi_1' + \cos g\phi_2' + \cdots + \cos g\phi_N'\right)\right\}^{2n}\right\rangle\nonumber\\
&=& m'^{4n}\sum_{r_1,r_2,\cdots ,r_N}^n\frac{\delta_{n,r_1+r_2+\cdots +r_N}}{(2r_1)!(2r_2)!\cdots(2r_N)!}\nonumber\\
&\times&\left\langle\left(\int dx\cos g\phi_1'\right)^{2r_1}\left(\int dx\cos g\phi_2'\right)^{2r_2}\cdots\left(\int dx\cos g\phi_N'\right)^{2r_N} \right\rangle\nonumber\\
&=& \left(\frac{m'^{2}}{2}\right)^{2n}\sum_{r_1,r_2,\cdots r_N}^n
\frac{\delta_{n,r_1+r_2+\cdots +r_N}}{(r_1!)^2(r_2!)^2 \cdots (r_N!)^2}\int\prod dxdy \nonumber\\
&\times &\left\langle\exp ig\left[\sum_i^{r_1}\{\phi_1'(x_{1,i})-\phi_1'(y_{1,i})\}+\cdots +\sum_i^{r_N}\{\phi_N'(x_{N,i})-\phi_N'(y_{N,i})\}\right]\right\rangle.\nonumber\\
&=& \left(\frac{m'^2}{2}\right)^{2n}\nonumber\\
&\times&\sum_{r_1,\cdots,r_N}^n\frac{\delta_{n,r_1+\cdots+r_N}}{(r_1!)^2\cdots (r_N!)^2}\int\prod dxdy\frac{\prod_{i>j}^{r_1}(x_{1,i}-x_{1,j})^2\mu^2(y_{1,i}-y_{1,j})^2\mu^2}{\prod_{i,j}^{r_1}(x_{1,i}-y_{1,j})^2\mu^2}\times\cdots\nonumber\\
&\times& \left\langle \exp ig\left[\sum_{i=1}^{r_1}\{\phi_1(x_{1,i})-\phi_1(y_{1,i})\}+\cdots + \sum_{i=1}^{r_N}\{\phi_N(x_{N,i})-\phi_N(y_{N,i})\}\right]\right\rangle\nonumber\\
&\times& \exp ng^2\left\{<\phi_i(0)^2>-<\phi_i'(0)^2>\right\}.\label{Z_B}
\end{eqnarray}
If we choose the renormalization of mass as
\begin{equation}
 m'^2=\frac{m\mu}{\pi}\exp \frac{g^2}{2}\left\{<\phi_i'^2>-<\phi_i^2>\right\},
\end{equation}
then Eq.(\ref{Z_B}) is equal to (\ref{Z_2n_cal}). 
From the free boson propagator (\ref{b'_prop}) and the potential term
(\ref{b_potential}) we find that the boson model with the Lagrangian 
\begin{equation}
{\cal L}_{\rm B}= \frac{g^2}{4(3g-2\pi)}\sum_{i,j}\left[\frac{g}{3g(N-1)+2\pi}+\frac{g-\pi}{\pi}\delta_{i,j}\right]\partial_{\mu}\phi_i'\partial^{\mu}\phi_j'
+ m'^2\sum_{i}\cos g\phi_i' \label{b_lagrangian}
\end{equation}
is equivalent to the fermion model of (\ref{lagrangian1}).

%% file: discussion.tex
\section{Discussion}

Let us calculate the eigenvalues of the coefficient matrix $\{A_{i,j}\}$ of the
kinetic term in the boson Lagrangian (\ref{b_lagrangian}), which is given by
\begin{equation}
 A_{i,j}= \frac{g^2}{2(3g-2\pi)}\left[\frac{g}{3g(N-1)+2\pi}+\frac{g-\pi}{\pi}\delta_{i,j}\right].
\end{equation}
From the equation
\begin{equation}
 {\rm det}\mid A_{i,j}-\alpha\delta_{i,j}\mid = 0
\end{equation}
we obtain
\begin{eqnarray}
 \alpha &=& \frac{g^2\{(N-1)g+\pi\}}{2\pi\{3g(N-1)+2\pi\}}\ \ :{\rm single\  root,}\label{single_eigen} \\
\alpha &=& \frac{g^2(g-\pi)}{2\pi(3g-2\pi)}\ \ :(N-1){\rm -degenerated\  roots}.\label{degene_eigen}
\end{eqnarray}
In order for the boson model (\ref{b_lagrangian}) to be physically
sensible, these eigenvalues must be non-negative, so we have
\begin{equation}
g < - \frac{\pi}{N-1}, \label{region_1}
\end{equation}
or 
\begin{equation}
 - \frac{2\pi}{3(N-1)}\leq g \leq\frac{2\pi}{3},\label{region_2}
\end{equation}
or 
\begin{equation}
 g> \pi.\label{region_3}
\end{equation}
It is hard to find such regions (\ref{region_1}), (\ref{region_2}) or
(\ref{region_3}) from direct observation of the fermion model
(\ref{lagrangian1}). In the previous paper I we missed the region
corresponding to (\ref{region_2}). It seems that even in the outside of
the regions (\ref{region_1}), (\ref{region_2}) and (\ref{region_3}), we
can change the sign of the kinetic part in the Lagrangian by
transforming the fields into imaginary ones. By such transformation,
however, the potential terms become  unbounded from blow. This means
that in order to fix the physical regions  of the coupling constant we
must check both kinetic and potential terms in the Lagrangian. The
kinetic terms in the Lagrangian (3.13) of the previous paper I look like
negative in the region (\ref{region_2}), but by scaling the boson field
we can still define a physically sensible boson model in this region;
see below. 

At the specific values
$g=-\frac{2\pi}{3(N-1)}$ or $g=\frac{2\pi}{3}$ the eigenvalues
(\ref{single_eigen}) or (\ref{degene_eigen})
diverge. When $g=-\frac{2\pi}{3(N-1)}$ one boson field corresponding to
(\ref{single_eigen}) disappears from the system and the original
$N$-species fermion model is described by $(N-1)$-species boson fields,
while for $g=\frac{2\pi}{3}$, $(N-1)$-boson fields corresponding to
(\ref{degene_eigen}) disappear and the original fermion model is
equivalent to the boson model with only one species, i.e. the ordinary
sine-Gordon model. 

For the case of $N=2$ the Lagrangian (\ref{b_lagrangian}) in the
previous section is rewritten as
\begin{eqnarray}
 {\cal L}_{\bf B} &=& \frac{g^2}{4(3g-2\pi)}\sum_{i,j}^2\left[\frac{g}{3g+2\pi}+\frac{g-\pi}{\pi}\delta_{i,j}\right]\partial_{\mu}\phi_i'\partial^{\mu}\phi_j'+m'^2(\cos g\phi_1' + \cos g\phi_2')\nonumber\\
&=& \frac{g^2}{4(3g-2\pi)}\left[\frac{g}{3g+2\pi}\left\{(\partial\phi_1')^2 + (\partial\phi_2')^2 + 2\partial\phi_1'\partial\phi_2'\right\}+\frac{g-\pi}{\pi}\left\{(\partial\phi_1')^2+(\partial\phi_2')^2\right\}\right]\nonumber\\
&&+2m'^2\cos\frac{g}{2}(\phi_1'+\phi_2')\cos\frac{g}{2}(\phi_1'-\phi_2').
\end{eqnarray} 
We put 
\begin{equation}
 \phi_{\pm}=\frac{g}{2}\sqrt{\frac{g\pm\pi}{\pi(3g\pm 2\pi)}}(\phi_1'\pm \phi_2')
\end{equation}
to obtain
\begin{eqnarray}
 {\cal L}_{\rm B}&=& \frac{1}{2}\left[(\partial\phi_+)^2+(
\partial\phi_-)^2\right]\nonumber\\
&&+2m'^2\cos\left(\sqrt{\frac{\pi(3g+2\pi)}{g+\pi}}\phi_+\right)\cos\left(\sqrt{\frac{\pi(3g-2\pi)}{g-\pi}}\phi_-\right),\label{two_boson}
\end{eqnarray}
which corresponds to the boson Lagrangian in I. As
mentioned before, it is seen that the Lagrangian above is invariant for
$g\rightarrow -g$ with $\phi_+ \leftrightarrow \phi_-$. There does
not exist, however, such a symmetry for the Lagrangian
(\ref{b_lagrangian}) when $N\geq 3$. 

If $\pi>g>2\pi/3$ or
$-2\pi/3>g>-\pi$, the potential term in (\ref{two_boson}) becomes 
\begin{equation}
 2m'^2\cosh\left(\sqrt{\frac{\pi (3g+2\pi)}{-g-\pi}}\phi_+\right)\cosh\left(\sqrt{\frac{\pi(3g-2\pi)}{-g+\pi}}\phi_-\right),
\end{equation}
which is unbounded from below, and the model is physically nonsensical. Then $g$
must be set in the region (\ref{region_1}), (\ref{region_2}) or
(\ref{region_3}) with $N=2$.  

In our path-integral formulation the massive Thirring-like model is described 
as the system of  massless fermions interacting with massless bosons
where the interaction part contains mass parameter. While the
bosonization technique may be applicable only for the charge-zero sectors,
it is an interesting subject to study the charged sectors, i.e. the
bound states of some particles in our formulation.